\documentclass{appolb}
\usepackage{bm}
\usepackage{epsfig}

\newif\ifpdf
\ifx\pdfoutput\undefined
\pdffalse 
\else
\pdfoutput=1 
\pdftrue
\fi

\def\OMIT#1{}

\newcommand{\bea}{\begin{eqnarray}}
\newcommand{\eea}{\end{eqnarray}}

\newcommand{\jpsi}{J/\psi}

\begin{document}
\date{\today}
\pagestyle{plain}
\newcount\eLiNe\eLiNe=\inputlineno\advance\eLiNe by -1

\ifpdf
\DeclareGraphicsExtensions{.pdf, .jpg}
\newcommand{\picspace}{\vspace{-2.5in}}
\newcommand{\picspacehalf}{\vspace{-1.75in}}
\else
\DeclareGraphicsExtensions{.eps, .jpg,.ps}
\newcommand{\picspace}{\vspace{0in}}
\newcommand{\picspacehalf}{\vspace{0in}}
\fi

\title{Recent Developments in Heavy Quark and Quarkonium Production%
\thanks{Presented at the XXXIII International Symposium on Multiparticle
Dynamics, September 5-11, 2003, Krakow, Poland}%
}
\author{Thomas Mehen
\address{Duke University}
}
\maketitle
\begin{abstract}

Recent measurements of $\jpsi$ production in $e^+e^-$ colliders pose a challenge to the NRQCD factorization theorem for
quarkonium production. Discrepancies between leading order calculations of color-octet contributions and the  momentum
distribution of $\jpsi$ observed by Belle and BaBar are resolved by resumming large perturbative and  nonperturbative
corrections that are enhanced near the kinematic endpoint. The large cross sections for $\jpsi c \bar c $ and double
quarkonium production remain poorly understood. Nonperturbative effects in fixed-target hadroproduction of open charm
are also discussed. Large asymmetries in the production of charm mesons and baryons probe nonperturbative corrections
to the QCD factorization theorem. A power correction called heavy-quark recombination can economically explain these
asymmetries with a few universal parameters.

\end{abstract} 


\section{Introduction} 

In the last couple of years there have been a number of interesting experimental  results in the production of heavy
particles, including  measurements of $J/\psi$ production in $e^+e^-$ collisions at the $\Upsilon(4S)$
resonance~\cite{Abe:2002rb,babar,belle} and charm meson production at the  Tevatron~\cite{Acosta:2003ax}.  This talk
focuses on how these results impact theoretical understanding of heavy particle production. The spectrum of $J/\psi$ 
observed in $e^+e^-$ colliders disagrees with leading order calculations based on Non-Relativistic QCD (NRQCD)
factorization theorems~\cite{nrqcd}.  Better agreement is obtained in calculations which resum the large
nonperturbative  and perturbative corrections that arise  near the kinematic endpoint~\cite{flm}. However, large cross
sections for $J/\psi c \bar c$ and exclusive double quarkonium production remain poorly understood. I also discuss
heavy quark production. Calculations of charm and bottom production which resum large logarithms of $p_\perp/m_Q$
provide a consistent description of the production of these particles at the
Tevatron~\cite{Binnewies:1998vm,Cacciari:2002pa,Cacciari:2003zu}. Finally, I explain how the large asymmetries observed
in fixed-target hadroproduction experiments can be explained by a power correction to the QCD factorization theorem
called heavy-quark recombination~\cite{bjm,Braaten:2003vy}.

\section{$J/\psi$ Production at the $\Upsilon(4S)$}

The current theoretical framework for understanding the production of heavy quarkonia is NRQCD~\cite{nrqcd}. NRQCD
solves important theoretical and phenomenological problems in quarkonium theory. Color-singlet model calculations of
$\chi_c$ decay suffer from infrared divergences \cite{Barbieri:1976fp}. NRQCD provides a generalized factorization
theorem that includes nonperturbative corrections to the color-singlet model, including color-octet decay and
production mechanisms. Infrared divergences are factored into nonperturbative matrix elements, so calculations of
inclusive production and decay rates are infrared safe~\cite{Bodwin:1992ye}. Color-octet production mechanisms are
necessary for understanding the production of $J/\psi$ at large transverse momentum, $p_\perp$, at the Tevatron
\cite{tevatron}. Convincing evidence for color-octet mechanisms has recently been seen in $\gamma \gamma$ collisions,
where color-singlet mechanisms underestimate the cross section by an order of magnitude while calculations including
color-octet mechanisms describe the data well~\cite{Klasen:2001cu}.

However, there are  many unsolved problems in quarkonium physics~\cite{Bodwin:2002mr}. Perhaps the most puzzling is the
polarization of $J/\psi$ and $\psi^\prime$, which is predicted to be transverse at very large $p_\perp$ in hadron
colliders~\cite{psipol}. The theoretical prediction is consistent with the data at intermediate  $p_\perp$ but at the
largest  $p_\perp$ measured the $J/\psi$ and $\psi^\prime$ are observed to be slightly longitudinally
polarized. 

This talk focuses on new puzzles arising from recent measurements of  $J/\psi$ production in  $e^+e^-$
colliders~\cite{Abe:2002rb,babar,belle}. 
The leading color-octet contribution to this process 
was expected to dramatically enhance the cross section for maximally energetic $\jpsi$  
as well as modify their angular distribution~\cite{bc}.
If $\cos \theta$ is the angle of the $\jpsi$ relative to the axis defined by the $e^+e^-$ beams and 
$z= E_{J/\psi}/E^{\rm
max}_{J/\psi}$, then the differential cross section is
\bea
\frac{d \sigma}{d z \, d \cos \theta} = S(z)(1+A(z)\cos^2 \theta) \, .
\eea
The function $A(z)$ tends to -1 as $z \to 1$ for color-singlet
production. The leading color-octet diagram contributes only at $z =1$
and gives $A(1) \approx 1$. The total color-singlet
cross section is predicted to be 0.4-0.9 pb~\cite{cs}, while the total cross section from the leading
color-octet mechanism is expected to be $\approx 1$ pb. A substantial rise in the cross section
near the kinematic endpoint accompanied by a change in angular distribution was predicted to be a robust signal of the 
color-octet mechanism~\cite{bc}. 

Experimental data does not agree with these expectations~\cite{Abe:2002rb,babar,belle}. One problem is that a sharp
rise in the cross section near the kinematic endpoint is not observed.  On the other hand, the cross section is larger
than predicted by the color-singlet model and $A(z) \approx 1$ for $0.7 < z < 1$.  Another puzzle  is the production of
open charm with $J/\psi$. Belle  observes that a large fraction of $J/\psi$ are produced with
an additional  $c \bar{c}$~\cite{Abe:2002rb}: 
\bea
\frac{\sigma (e^+e^- \rightarrow J/\psi \, c \bar{c})}{\sigma (e^+e^- \rightarrow J/\psi \,X)}
=0.59^{+0.15}_{-0.13}\pm 0.12.
\eea 
(Preliminary results described by P. Pakhlov in this conference
suggest that the ratio  is even larger~\cite{pp}.) Leading order color-singlet model calculations predict the ratio to be $\approx
0.2$~\cite{cs} and a large color-octet contribution will make the ratio even smaller. 

The resolution of the first problem lies in a careful analysis of the perturbative and nonperturbative corrections that appear
near the kinematic endpoint of quarkonia production \cite{flm,brw}.  The 
$J/\psi$ production cross section in NRQCD is
\begin{equation} \label{NRQCDprod} 
\frac{d \sigma}{dz} (e^+ e^- \rightarrow \jpsi + X) = \sum_n \frac{d \hat \sigma}{dz} (e^+ e^- \rightarrow c
\bar{c}[n]+ X)
\langle {\cal O}^{\jpsi}_n \rangle \,,
\end{equation}
where $\langle {\cal O}^{\jpsi}_n \rangle$ are NRQCD matrix elements and $d\hat{\sigma} (e^+ e^- \rightarrow c\bar{c}[n]+ X)$
are perturbatively calculable short-distance cross sections. The label $n$ denotes the angular momentum and 
color quantum numbers of the $c \bar c$. The NRQCD scaling rules show that $\langle {\cal O}^{\jpsi}_n \rangle$
scales as some power of $v$, where $v$ is the typical velocity of the heavy quarks inside the bound state.

The leading Feynman diagrams which  contribute to color-octet  production  give $d\hat{\sigma} (e^+ e^- \rightarrow c
\bar{c}[n]+ X)/dz$  proportional to $\delta(1-z)$. This is the first in an infinite series of terms that are singular at $z = 1$.
There are  higher order nonperturbative corrections that  scale as $v^{2n}/(1-z)^n$ \cite{brw} as well as 
perturbative corrections that go like $\alpha_s^n {\rm ln}^m(1-z)/(1-z)$, $m \leq 2 n-1$. For charmonium, $v^2 \sim \alpha_s
\sim 0.3$ so for $z \geq 0.7$  perturbation theory and the NRQCD $v$ expansion both break down.

Near the kinematic endpoint, the final state consists of two kinds of quanta: energetic collinear particles with
light-like momenta in the jet against which the $J/\psi$ is recoiling and particles which are soft as
viewed from the rest frame of the $J/\psi$. NRQCD breaks down because the theory does not explicitly include these
degrees of freedom.     The problem can be fixed by  using  the Soft-Collinear Effective Theory (SCET)~\cite{scet}
which explicitly includes both collinear and soft degrees of freedom. By combining SCET and NRQCD one finds that in the
endpoint region  Eq.~(\ref{NRQCDprod}) is replaced with  the following factorization theorem:~\cite{flm}
\begin{equation}\label{factformula}
\frac{d\sigma^{[n]}}{dz} =   \sigma_0^{[n]}
\int_z^{1} d\xi \, S^{[n]}(\xi) \, J(\xi-z)\, .
\end{equation}
Here $\sigma_0^{[n]}$ is a short distance cross section which is perturbatively calculable. The shape 
function $S^{[n]}(\xi)$ is a universal
nonperturbative distribution that resums the large  nonperturbative  corrections.  $J(\xi-z)$ is a
calculable function that describes the propagation of the collinear particles in the energetic  gluon jet. Large perturbative
corrections can be  resummed by solving the SCET renormalization group equations for $\sigma_0^{[n]}, S^{[n]}(\xi)$ and
$J(\xi-z)$. 
\begin{figure}
\begin{center}
 \includegraphics[width=5 in]{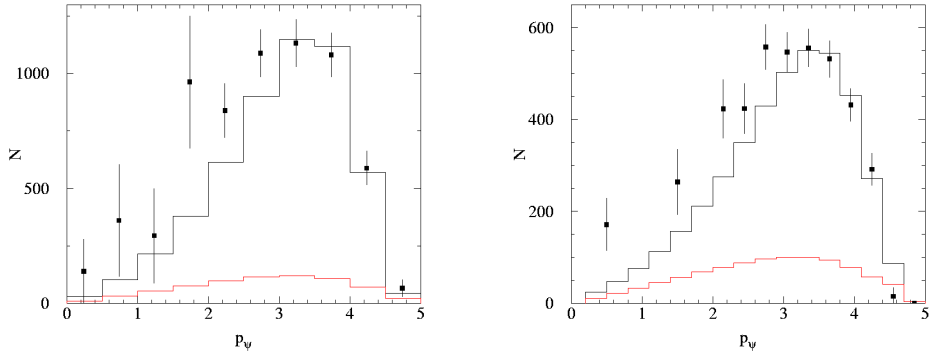}
\end{center}
\caption{\it  The sum of the resummed color-octet and leading order color-singlet contributions are plotted as the upper line. 
The lower line is the leading order color-singlet contribution only, and the data are from the BaBar collaboration ~\cite{babar} (left)
and  Belle collaboration~\cite{belle} (right).} 
\label{comp}
\end{figure}

Comparison of this calculation with data from BaBar~\cite{babar} and Belle~\cite{belle} is shown in
Fig.~\ref{comp}~\cite{flm}. The number of events is plotted as a function of the $J/\psi$ momentum,
$p_\psi$. The lower line in both plots is the leading color-singlet contribution which falls well
below data.  The upper line includes the resummed color-octet cross section. The leading order
calculation of the color-octet contribution (not shown) gives an integrated cross section comparable
to that of the resummed calculation shown here, but the entire cross section is located in the very
last bin of $p_\psi$. This also does not agree with data. Fig.~\ref{comp} demonstrates that when
large perturbative and nonperturbative corrections are included the momentum spectrum of the
$J/\psi$  produced by the color-octet mechanism is significantly broadened. The agreement with data
is good in the endpoint region where the calculation is most reliable.  The calculation is not
predictive because the shape function is fitted to available data. However, the moments of the shape
function  can be estimated using the NRQCD scaling rules  and the shape function used in
Ref.~\cite{flm} satisfies these constraints. The universality  of the shape function can be tested by
applying the methods of Ref.~\cite{flm}   to other $J/\psi$ production processes.

While resolving the discrepancy between leading order color-octet calculations and the observed $p_\psi$ distribution,
the calculation does not help explain the large cross section for $\jpsi c \bar c$ observed by the Belle collaboration.
The Belle collaboration also observes a large cross section for exclusive $J/\psi + \eta_c$ and $J/\psi +
\chi_c$~\cite{belle}. An NRQCD analysis of exclusive cross sections appears in Ref.~\cite{Braaten:2002fi}.
Because the helicity conservation rules for exclusive QCD processes suppress the leading QCD contribution, the purely
QED contributions are surprisingly large ($\approx$ 20\%). The leading relativistic corrections give large corrections which
unfortunately are difficult to estimate reliably. For instance, Ref.~\cite{Braaten:2002fi} quotes
$\sigma(J/\psi+\eta_c) = 5.5^{+10.6}_{-3.5}$ fb, with the uncertainty dominated by relativistic corrections. For
comparison Belle measures $\sigma(J/\psi+\eta_c) \times{\rm Br}[\eta_c \to {\rm 4 \,charged \,
particles}] = 33^{+7}_{-6}\pm 9$ fb. Clearly a mechanism  for enhancing the $J/\psi c \bar c$ and double charmonium
cross sections is needed. Proposals for nonperturbative mechanisms responsible for this enhancement appear in
Ref.~\cite{spec}.

\section{Open Charm and Bottom Production}

The QCD factorization theorem states that the production cross
section for a heavy particle $H$ containing a heavy quark $Q$     is \cite{css}
\begin{eqnarray}\label{fact}
d \sigma [A B &\rightarrow & H X] =  \sum_{i,j} f_{i/A} \otimes f_{j/B} \otimes d\hat{\sigma}[ij \rightarrow Q \bar Q X] \otimes
D_{Q \rightarrow H} + ... \, . 
\end{eqnarray}
Here $f_{i/A}$ is a parton distribution function, $D_{Q \rightarrow H}$ is a fragmentation function, $d\hat{\sigma}[ij
\rightarrow Q\bar Q X]$ is a  short-distance cross section and the ellipsis represents corrections which are
suppressed by $\Lambda_{\rm QCD}/m_Q$ or $\Lambda_{\rm QCD}/p_\perp$. 

Perturbative aspects of Eq.~(\ref{fact}) are tested by measurements of charm and bottom production at 
collider experiments. Experimental reviews of heavy quark production at LEP,  HERA and  the Tevatron are described in the talks of 
A.~Sciaba~\cite{Sciaba}, B.~Olivier~\cite{Olivier} and K.~Rinnert~\cite{Rinnert}, respectively. 
At the Tevatron, discrepancies between NLO calculations of bottom production and experimental cross sections have been
known for a long time~\cite{Acosta:2001rz}. Recently CDF has extended measurements to include charm as well
as bottom~\cite{Acosta:2003ax}. Resummation of large logarithms of $p_\perp/m_Q$ is needed to obtain better agreement
with both charm and bottom cross sections~\cite{Binnewies:1998vm,Cacciari:2002pa,Cacciari:2003zu}. It is also important
to treat fragmentation functions carefully as the total cross section is sensitive to the fragmentation function
used~\cite{Cacciari:2003zu}. Though the existing calculations differ in how finite mass corrections are handled they
are consistent numerically and agree with data within the theoretical uncertainties estimated by varying factorization
and renormalization scales. 

Nonperturbative power corrections to Eq.~(\ref{fact}) are probed in lower energy fixed-target experiments. Production asymmetries are an
incisive test of these corrections. At leading order in perturbation theory, charm particles and antiparticles are produced symmetrically
because the  partonic processes $gg\rightarrow c\bar c$ and $q \bar q \rightarrow c \bar c$ produce charm  and anticharm  symmetrically and 
$D_{c\rightarrow H} = D_{\overline c \rightarrow \overline H}$ due to charge conjugation invariance.  At next-to-leading order, the asymmetry,
$\alpha[H] = (\sigma[H] - \sigma[\,\overline H \,])/ (\sigma[H] + \sigma[\,\overline H \,])$, is only a few percent~\cite{nlo}. Fixed-target
hadroproduction \cite{hadro,Aitala:2000rd,E791,selex} and photoproduction \cite{photo,focus:2003xh} experiments observe much larger 
asymmetries.  In hadroproduction the asymmetries are known as the ``Leading Particle Effect''. Cross sections for charm particles sharing a
valence quark with the beam hadron are enhanced in the forward direction of the beam. Hadroproduction asymmetries can be quite large. For
example, in the most forward region measured in $\pi^- N$ collisions, the ratio of $D^-$ to $D^+$ is $\approx 6$.  

Charm asymmetries are conventionally explained by nonperturbative models of hadronization \cite{models}.  These models suffer
from a lack of predictive power, since they depend on a number of nonperturbative functions, such as the distribution of
spectator quarks in hadron remnants. The most  commonly used model is the Lund string fragmentation
model~\cite{Sjostrand:1986ys}  which can be implemented using PYTHIA~\cite{Sjostrand:2001yu}.  The PYTHIA Monte Carlo with
default parameters rarely  predicts the asymmetries correctly~\cite{hadro} and in the case of $\Lambda_c$ asymmetries in $\pi
N$ collisions~\cite{Aitala:2000rd} and $\gamma N$ collisions \cite{focus:2003xh}  gets the sign of the asymmetry wrong.  

A novel mechanism for generating charm hadron asymmetries  called heavy-quark recombination has recently been introduced in
Ref.~\cite{bjm}.  Similar mechanisms for production of light hadrons were considered in Ref. \cite{recom}.  An important
difference between the production of heavy hadrons and light hadrons is that in the former case heavy quark symmetry \cite{iw}
can be used to simplify the structure of nonperturbative factors appearing in the calculation.  Heavy-quark recombination is an
$O(\Lambda_{\rm QCD}/m_c)$ suppressed power  correction to the factorization theorem of Eq.~(\ref{fact}). In this process, a
light anti-quark, $\overline{q}$, from the incident hadron participates in a hard-scattering process  which produces a $c$ and
$\overline{c}$ quark. Following the hard scattering the  $\overline{q}$ and the $c$ recombine to form a $D$ meson. A similar
mechanism in which  a light quark recombines with the $c$ quark is the dominant recombination contribution
to charm baryon production \cite{Braaten:2003vy}. Heavy-quark recombination differs from previous nonperturbative models in
that the asymmetry is generated in the short-distance process so cross  sections are calculable up to an overall normalization
that is set by a few universal parameters.  The short-distance cross section is strongly  peaked in the forward direction of
the light quark or antiquark,  naturally leading to an asymmetric cross section. 
\begin{figure}
\begin{center}
 \includegraphics[width=5 in]{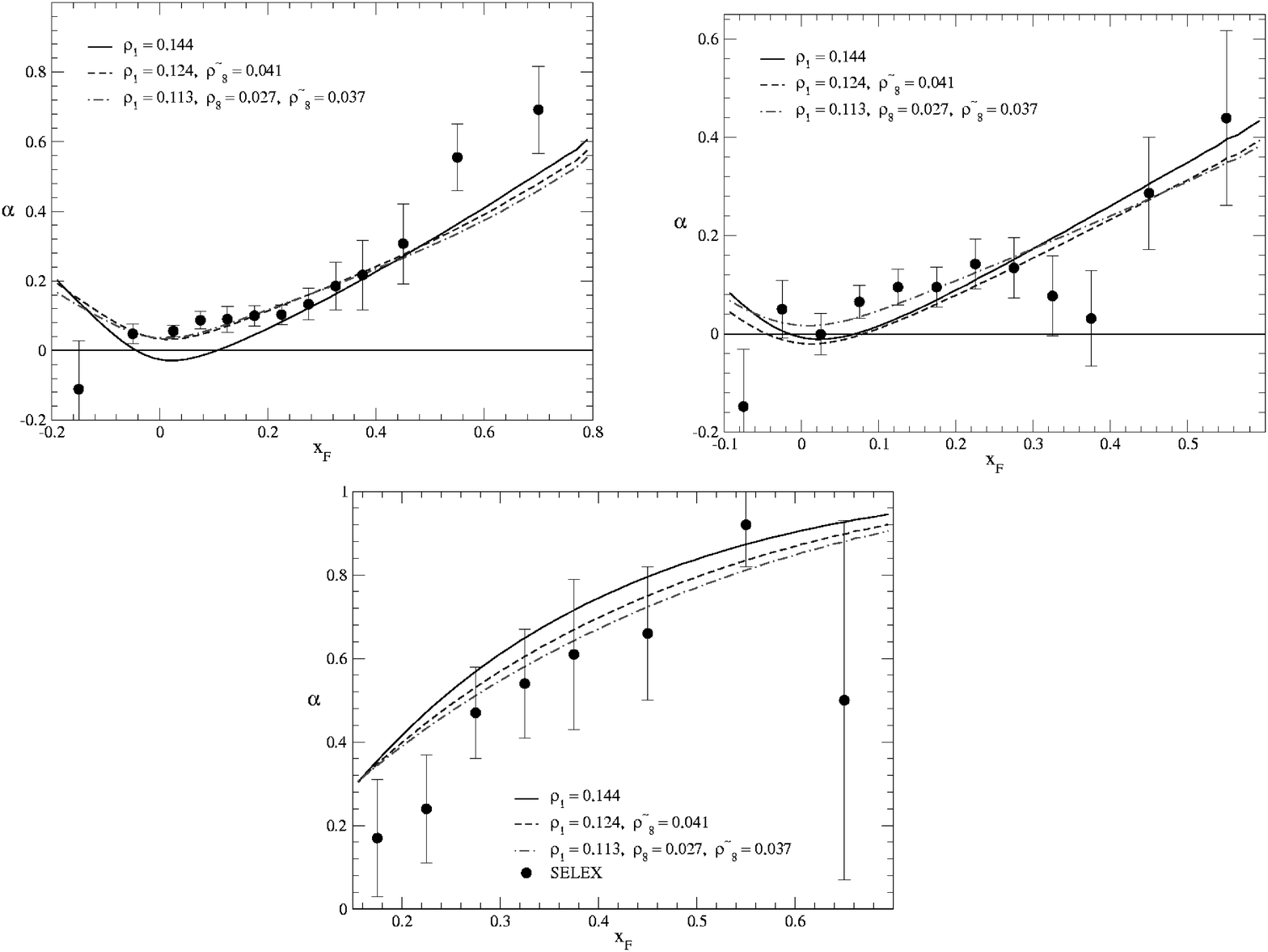}
\end{center}
\caption{\it  Asymmetry as a function of $x_F$ for $D^+$ mesons (top left) and  $D^{*+}$ mesons (top right)
produced in 500 GeV $\pi N$ collisions \cite{E791} and
$D_s^+$ mesons produced in 600 GeV $\Sigma^- N$ collisions  \cite{selex} (bottom). Theory curves are 
explained in text.} 
\label{univ}
\end{figure}
 
Heavy-quark recombination accounts for the  $D$ meson asymmetries observed in photoproduction and hadroproduction experiments.
Fig.~\ref{univ} shows asymmetries for $D^+$ and $D^{*+}$ mesons produced in 500 GeV $\pi N$  collisions \cite{E791} and asymmetries for
$D_s$ mesons produced in 600 GeV $\Sigma^- N$ collisions  \cite{selex}. There are four universal parameters in the theory. The theory
curves in Fig.~\ref{univ} correspond to fits with one, two and three of these parameters. (A four parameter fit yields identical results
as the three parameter fit.) The heavy-quark recombination mechanism correctly describes asymmetries for different types of $D$ mesons in
experiments with different beams with a minimal set of universal parameters. The heavy-quark recombination
also correctly describes $\Lambda_c$ asymmetries in both $\pi N$ and $p N$ collisions~\cite{Braaten:2003vy}.

\end{document}